\documentclass[twocolumn,letter]{jpsj3}
\usepackage{txfonts,bm,color,ulem}
\newcommand{\1}{\mbox{1}\hspace{-0.25em}\mbox{l}}
\title{
Symmetry Protected Weak Topological Phases 
in a Superlattice 
}

\author{
Takahiro {\sc Fukui},
Ken-Ichiro {\sc Imura}$^1$, and 
Yasuhiro {\sc Hatsugai}
$^2$
}
\inst{Department of Physics, Ibaraki University, Mito 310-8512, Japan\\
$^1$Department of Quantum Matter, AdSM, Hiroshima University, Higashi-Hiroshima 739-8530, Japan\\
$^2$Institute of Physics, University of Tsukuba, 1-1-1 Tennodai, Tsukuba, Ibaraki 305-8571, Japan}

\recdate{ \hspace{50mm} }

\abst{
We explore novel topological phases 
realized in a superlattice system based on the Wilson-Dirac model.
Our main focus is on a two-dimensional analogue of weak topological insulator phases.
We find such phases as those
characterized by gapless edge states that are protected by symmetry but
sensitive to the orientation of the edge relative to the superlattice structure.
We show that manifest and hidden reflection symmetries protect 
such  weak topological phases,
and propose bulk
$\mathbb{Z}_2$ indices responsible for the topological protection of the edge states.
}

\kword{topological insulator, superlattice, Chern number, 
weak topological phase, Wilson-Dirac model}

\begin{document}
\sloppy
\maketitle

Topological classification is a new trend in the field of condensed matter.
Although accepted only recently by the wide community in condensed matter physics, 
its position is very influential 
for determining future directions of the field.
Beyond the Ginzburg-Landau paradigm, topological classification can be applicable
to quantum liquids without fundamental symmetry breaking.\cite{wen89,Hatsugai06}
Still,  truly generic systems are not interesting and the symmetry again
restricts the systems.
Then, we have various physically interesting phases 
protected by symmetries.
The periodic table of topological phases 
as an extension of the classical 
symmetry classes is situated at the heart of the idea
\cite{Zirnbauer96,AltZir97,Kitaev08,SRFL08,HasKan10,QiZha11}.
Recently,
the extension of the standard classification scheme by including the diversity of topological phases
protected by other types of symmetry
has been investigated.\cite{Sato10,Chen11,Fid11,Pollmann12,TCI, Zaanen, Ai,CYR13}

Topological phases are mostly gapped; thus, their bulk is characterized by the
absence of low-energy excitations. On the other hand, with boundaries or impurities, 
there exist peculiar localized modes as edge states. The emergence of the
edge states is not accidental and is a fundamental property of topological phases, 
known as the bulk-edge correspondence\cite{Hatsugai93,HasKan10,QiZha11}. 
The edge states further reflect 
symmetries of topological phases and describe their variety beyond the
bulk characterization.

In this study, we attempt to further extend the idea of 
the topological classification
to a superlattice version of the Wilson-Dirac-type lattice model
that exhibits  hidden reflection symmetry.
The motivation of this work is not
purely academic.
Now that basic understanding of simple topological insulator crystals 
has been established,
a possible direction of not only theoretical research 
but also experimental research 
is to seek various topological quantum phenomena.
Recently, 
multilayer heterostructures 
consisting of alternating layers of topological and ordinary insulators
have been experimentally realized,
exhibiting an interesting correlation of bulk and surface properties.
\cite{Nakayama12,Valla12}

\begin{figure}
\begin{center}
%\begin{tabular}{cc}
\includegraphics[width=0.7\linewidth]{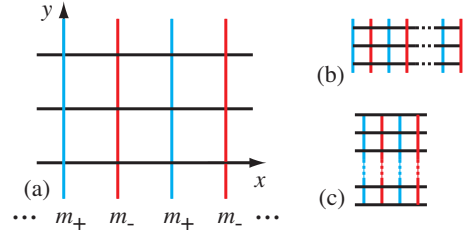}
%\end{tabular}
\caption{(Color online) Superlattice structure:
(a) schematic illustration of the lattice and
(b-c) different types of ribbon geometry.
The two edges of the ribbon are
along the $y$-axis in (b) and along the $x$-axis in (c).
}
\label{f:Model}%-----------------------------------------------
\end{center}
\end{figure}

\begin{figure}
\begin{center}
\includegraphics[width=0.6\linewidth]{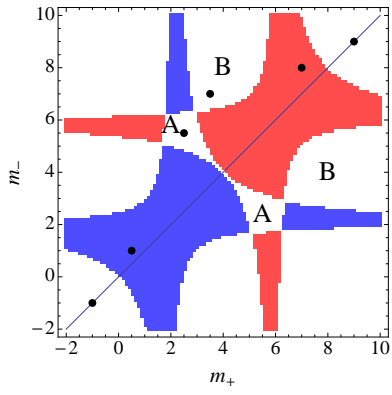}
\caption{(Color online) 
Phase diagram of the superlattice model for
$b=2$ and $t=1$:
Chern number $C$ evaluated at each point on the $(m_+, m_-)$-plane
is indicated by different colors:
blue, $C=1$; red, $C=-1$; and white, $C=0$.
Thick black dots correspond to the reference points in Table \ref{t:Inv}.}
\label{f:Pha}%-----------------------------------------------
\end{center}
\end{figure}

In this letter, we highlight a two-dimensional (2D) analogue of such
a superlattice system:
a variant of the Wilson-Dirac type tight-binding model 
with a stripe structure  (see Fig. \ref{f:Model}).
Without such a spatial nonuniformity, 
the simple 2D Wilson-Dirac model is known as the typical $\mathbb{Z}$-type model
specified by the Chern number \cite{QHZ08}.
The extension of this model to the quantum spin Hall effect (QSHE) \cite{KanMel05a} 
has been carried out \cite{BHZ},
in which the $\mathbb{Z}_2$ invariant \cite{KanMel05b,FuKan06} distinguishes 
between the QSHE and trivial phases.   
Here, we demonstrate that
the superlattice version exhibits a richer phase diagram (see Fig. \ref{f:Pha})
that cannot be classified by a single topological invariant.
Since the mass parameter $m$ controls the Chern number $c$ in a {\it uniform} system,
the superlattice of the mass $(m_+,m_-)$ can be regarded as that of distinct Chern insulators $(c_+,c_-)$.
On the other hand, this model has the total Chern number $C$ in its own right. 
We find that, in the  $C=0$ sector,
% of the vanishing Chern number
there appears a 2D analogue of weak topological insulating phases \cite{WTI1,WTI2,WTI3}
characterized by anisotropic topological properties.
Let us mention here the system's similarity to graphene.
Although graphene is already gapless in the bulk, it exhibits
direction-dependent boundary states
%Although it is gapless as a two dimensional system,
%we mention here similarity of graphene where there exist direction dependent
%boundary states 
with time reversal symmetry ($C=0$)\cite{Fujita96,RyuHat02,Hat09}. 

%%%%%%%%%%%%%%%%%%%%%%%%%%%%%%
%%%%%%%%%%%%%%%%%%%%%%%%%%%%%%

Figure \ref{f:Model} shows a schematic configuration of the superlattice model considered 
in this letter. 
%from now on.
%In real space the model can be specified by
%a tight-binding Hamiltonian 
%with only nearest neighbor hopping terms.
The model is defined on a square lattice, and on each site $\bm r = (x, y)$
of the lattice
an electron is allowed to occupy two orbital (pseudo-spin) states
with which 
a set of Pauli matrices $\sigma_\mu$ ($\mu=x, y, z$)
is associated.
The Hamiltonian is defined by 
\begin{alignat}1
H &= \sum_{\bm r} \left[\sum_{\mu=x,y} 
\left(
|\bm r \rangle \Gamma_\mu \langle \bm r + \hat{\bm \mu} | +
|\bm r + \hat{\bm \mu} \rangle \Gamma_\mu^\dagger \langle \bm r | 
\right) 
%\nonumber \\
+ 
%\sum_{\bm r} 
| \bm r \rangle V(\bm r)  \langle  \bm r |\right],
\label{Ham}
\end{alignat}
where 
$\hat{\bm \mu}$ stands for the unit vector in the $\mu$-direction, 
and the hopping and on-site potential terms 
%(each represented, respectively, by $\Gamma_\mu$ and $V(\bm r)$ in the equations below)
are respectively specified as
\begin{alignat}1
\Gamma_\mu &= -\frac{{\rm i}t}{2} \sigma_\mu + \frac{b}{2}\sigma_z,
\nonumber \\
V(\bm r) &= \left[m+(-1)^{x}\delta m - 2b\right] \sigma_z
\equiv \left(m_\pm - 2b\right) \sigma_z.
\end{alignat}
Note that  the lattice constant has been chosen to be unity; hence,
$x$ and $y$ take only integral values.
In this letter,
we focus on the simplest superlattice structure shown in Fig. \ref{f:Model}
in which
$V(\bm r)$ takes two alternating values
on each column of the vertical stripe.
%\cite{gen}
Taking into account
the doubling of the unit cell due to the stripe texture
as a sublattice degree of freedom,
one can block-diagonalize eq. (\ref{Ham})
in the reciprocal space as $H=\sum_{\bm k}|\bm k\rangle {\cal H}(\bm k)\langle \bm k|$ whose matrix element
${\cal H}(\bm k)$ is given by
\begin{equation}
{\cal H}(\bm k) =
\left(
\begin{array}{cc}
M_++ t\sin k_y \sigma_y& \Gamma_x + e^{-2{\rm i}k_x} \Gamma_x^\dagger\\
\Gamma_x^\dagger + e^{2{\rm i}k_x} \Gamma_x & M_- + t\sin k_y\sigma_y
\end{array}
\right),
\label{HamK}%----------
\end{equation}
where different rows and columns specify the sublattice,
{\it i.e.}, whether the electron is on a blue column ($x$: even) 
or  a red one ($x$: odd)
in Fig. \ref{f:Model}(a), and
$M_\pm = \left[m_\pm + b(\cos k_y-2)\right]\sigma_z$.
%We have also introduced
%$A_\mu =t \sin k_\mu\sigma_\mu$.

%%%%%%%%%%%%%%%%%%%%%%%%%%%%%%%%%%%%
%%%%%%%%%%%%%%%%%%%%%%%%%%%%%%%%%%%%

The uniform line $m_+ = m_-=m$ ($\delta m = 0$)
%the tight-binding Hamiltonian given as in either eq. (\ref{Ham}) or (\ref{HamK}) 
%reduces to that of 
corresponds to the standard Wilson-Dirac model\cite{BHZ,QHZ08}
%prescribed by 
${\cal H}(\bm k) = t\sum_\mu\sin k_\mu\sigma_\mu + m (\bm k)\sigma_z$,
where 
%$m (\bm k) = [m + b(\cos k_x + \cos k_y - 2)] \sigma_z$.
$m (\bm k) = \big[m + b\sum_\mu(\cos k_\mu -1)\big] \sigma_z$.
The half-filled ground state of the model
is classified by the Chern number\cite{TKNN,K_ann} $C$
that takes a nontrivial value of 
$C=1$ when $0<m/b<2$, while $C= -1$ when $2<m/b<4$
[otherwise, $C$ takes a trivial value ($C=0$)];
the change in the topological number corresponds to the closing of the gap
[zeros of $m (\bm k)$] at the Dirac point $\bm k=(0,0)$ and at its doublers' points 
%at either of the four inversion symmetric points:
%$\bm k = (0,0), 
$(\pi,0), (\pi, \pi), (0, \pi)$.
Away from the uniform line, 
%it is less obvious whether one can
%still classify different topological phases in terms of the same topological number $C$.
%Technically, 
it is still possible to compute the Chern number $C$ 
by applying the method given in ref. \cite{FHS05}
to the present superlattice model.
The phase diagram thus obtained is shown in Fig.~\ref{f:Pha}.

\begin{figure}
\begin{center}
\begin{tabular}{cc}
\includegraphics[width=0.45\linewidth]{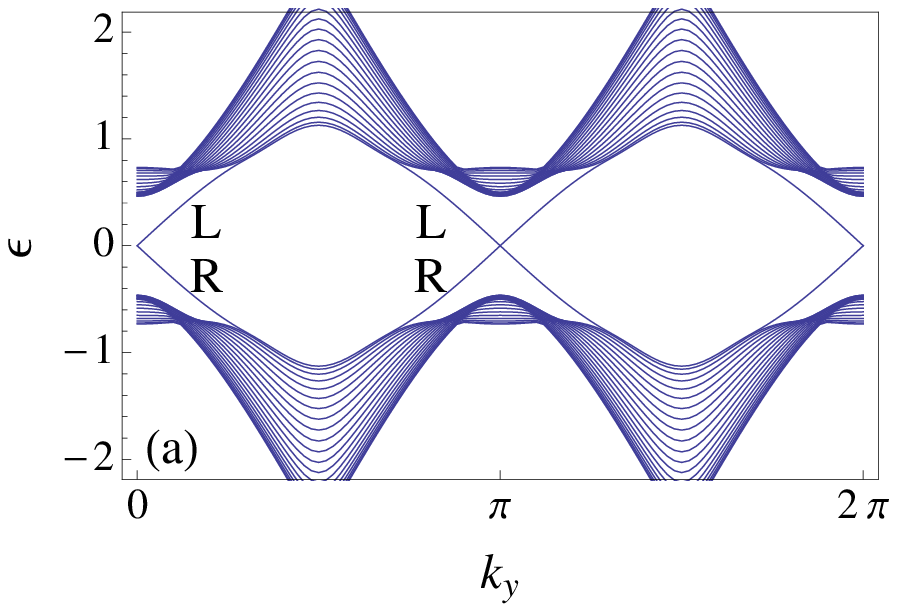}
&\includegraphics[width=0.45\linewidth]{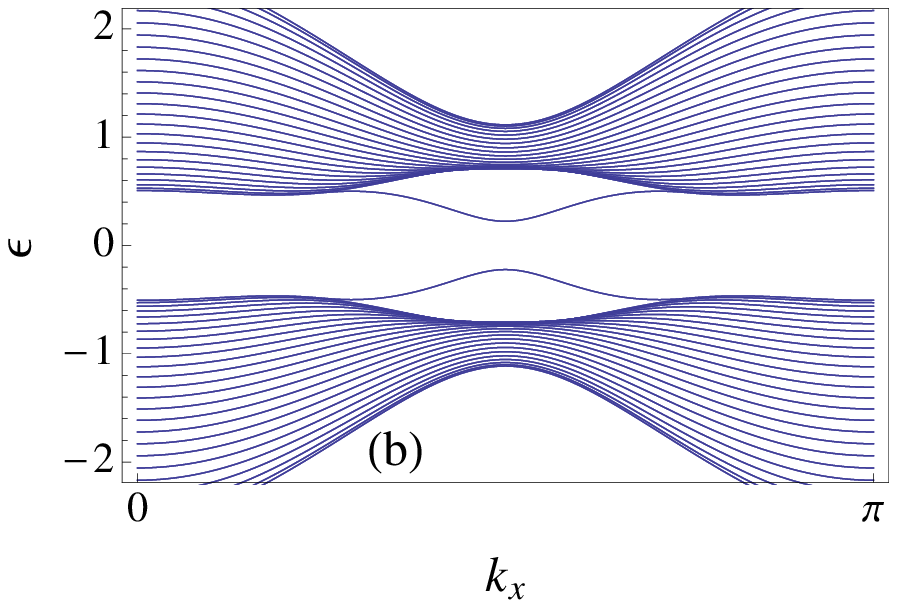}\\
\includegraphics[width=0.45\linewidth]{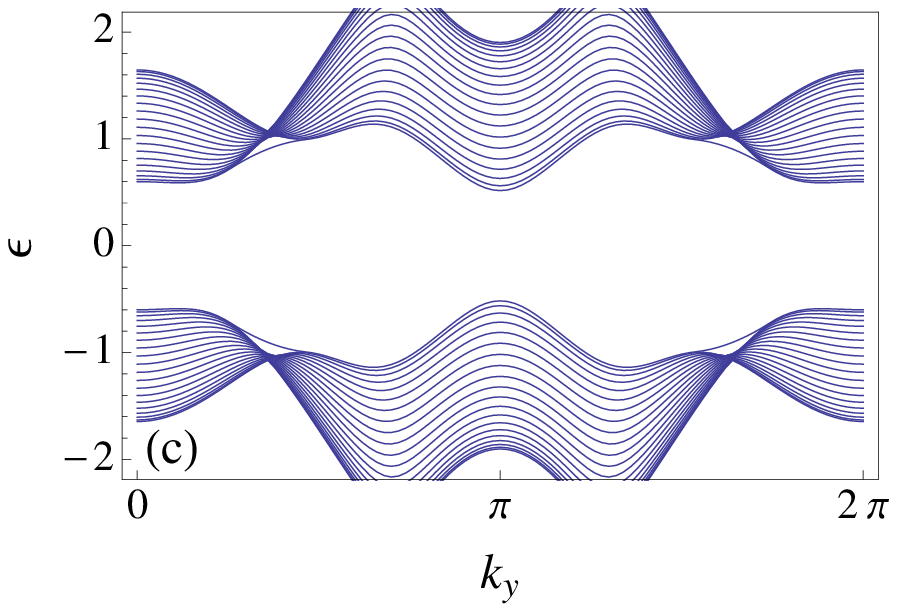}
&\includegraphics[width=0.45\linewidth]{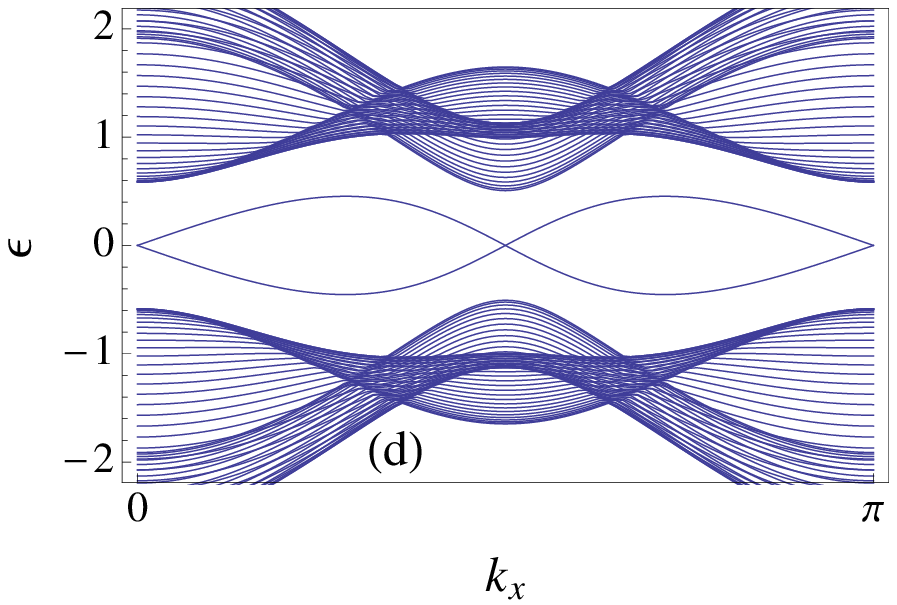}\\
\end{tabular}
\caption{(Color online) 
Energy spectrum in the ribbon geometry:
$b=2$, $t=1$ with boundaries along the $y$- (left) and 
$x$- (right) directions.
The upper [lower] panels: (a), (b) [(c), (d)] 
correspond to the case of mass parameters $m_+=2.5$, $m_-=5.5$ 
[$m_+=3.5$, $m_-=7$] belonging to phase A [B].
In (a), the suffixes L and R indicate that the corresponding edge state 
is localized at the left and right boundaries in Fig. \ref{f:Model}(b), respectively.
}
\label{f:edge_1}%-----------------------------------------------
\end{center}
\end{figure}

%The phase diagram of the superlattice model 
This phase diagram has the following specific features:
Topologically nontrivial phases with nonzero Chern numbers $C=\pm 1$
extend from the uniform line to a region of $m_+ \neq m_-$.
The regions of $C=\pm 1$ overlap (
at least they appear to do so in the phase diagram)
to form a finite domain of the $C=0$ phase
represented by A in Fig. \ref{f:Pha}.
There appear other $C=0$ phases in different parts of the phase diagram
separated by topologically nontrivial phases.
Are these $C=0$ phases simply topologically trivial?
Our answer is ``No''
%for example,
in the phases represented by ``A'' and ``B''.
%To our surprise
These phases exhibit gapless edge states in the ribbon geometry
[panels (a) and (d) of Fig.~\ref{f:edge_1}].
Interestingly,
the way these edge states appear depends on the way
the system's boundaries are introduced
(compare the top and bottom rows of Fig. \ref{f:edge_1}).
The structure of the phase diagram,
particularly the shape of  regions A and B is strongly dependent on
the hopping amplitude $t$
in contrast to the phase boundaries on the uniform line $m_+ = m_-$.
\cite{2scales}
It should be emphasized that, although
the concrete arrangement of distinct regions shown in Fig. \ref{f:Pha}
is not generic and varies continuously
as a function of $t$,
the behaviors of the edge states in the two types of $C=0$ phases
A and B are generic, implying that they are protected by some symmetry.

%%%%%%%%%%%%%%%%%%%%%%%%%%%%%%%%%
%%%%%%%%%%%%%%%%%%%%%%%%%%%%%%%%%

The concept of the bulk-edge correspondence
is now established,
\cite{Hatsugai93,Volovik03,EssGur11,FSFF12}
indicating that
the topological property of the bulk is fully reflected
in the spectrum of the edge states.
In the rest of this letter, we show 
that the gapless edge modes 
found in $C=0$ phases A and B 
are indeed topologically protected,
and interpret
the corresponding $C=0$ phases as a 2D analogue of 
the weak topological insulator in 3D.\cite{WTI1,WTI2,WTI3}
Figure \ref{f:edge_1} highlights the ``weak'' nature of phases A and B.

In phase A, edge states appear at the boundaries parallel to the $y$-axis
[Fig. \ref{f:edge_1}(a)],
whereas no edge states appear along those parallel to the $x$-axis [Fig. \ref{f:edge_1}(c)]. 
%The spectrum (a) reminds us of the QSHE of the graphene. \cite{KanMel05b} 
%However, in the latter TRS guarantees the level-crossing of the edge states, whereas in the present model, not only symmetries but also the momentum of the edge states at the crossing points are needed for the existence of the edge states.
%This is due to $(CP)\tilde{h}(\bm k)(CP)^{-1}=-\tilde{h}(\bm k)$ for the bulk and this reflects to the edge states.
As indicated in Fig. \ref{f:edge_1}(a)
a pair of counter-propagating modes,
one localized at the left boundary (L) and the other at the right boundary (R),
cross at zero energy
and at a specific momentum (symmetric point),
either at $k_y=0$ or $k_y=\pi$.
As we will show soon, this phase is reminiscent of the QSHE, since $C=c_++c_-=0$ for $c_+=1$ and $c_-=-1$,
and one pair of edge states is due to $c_+=1$, and the other pair is due to $c_-=-1$.
%In passing, we mention 
Note that, in the present model,
the existence of
particle-hole and inversion (or reflection) symmetries ensures
the spectrum at each $\bm k$ to be symmetric with respect to zero. 
This is also the case with the edge spectrum.

%In the phase diagram (Fig. 2)
%the phase A is restricted to a finite domain, while 
Phase B is, on the other hand, understood by considering the limit 
%extended to 
$m_-\rightarrow\infty$.
We claim that the midgap states in Fig. \ref{f:edge_1}(d) can be deformed
into topologically protected flat bands in this limit.
%without the bulk gap closing. 
%there appear in this limit zero energy flat bands which are
%the origin of the mid-gap states in Fig. \ref{f:edge_1} (d).
%can be understood
%in terms of the zero energy flat bands 
%%%%%%%%characteristic to systems with a chiral symmetry. \cite{RyuHat02}
To be concrete, when $m_-\rightarrow\infty$, electron occupation at $m_-$ sites is suppressed, 
and the model reduces to just a set of isolated one-dimensional ladders
% with a single mass parameter $m_+$
%%%%%%%%Its spectrum has no dispersion with respect to $k_x$,
%since hopping to the adjacent $m_-$ sites is suppressed.
%Neglecting the $m_-$ sites, the system becomes just a collection of isolated one-dimensional ladders
described by the reduced Hamiltonian
\begin{alignat}1
{\cal H}(k_y)=t\sin k_y\sigma_y+\left[m_+ + b(\cos k_y-2)\right]\sigma_z .
\end{alignat}
Owing to the chiral symmetry of this Hamiltonian,
the Berry phase integrated over $k_y$ is quantized to $0$ or $\pi$.\cite{RyuHat02,Hat09,STYY11}
To see this, set $Y=t\sin k_y$ and $Z=m_++b(\cos k_y-2)$.
Then, if the origin $(0,0)$ is located inside the ellipse $(Y,Z)$ forms in the $Y$-$Z$ plane,
i.e., $1<m_+/b<3$, the Berry phase is $\pi$ (nontrivial)
in which zero-energy flat bands are expected.\cite{RyuHat02,Hat09,STYY11}
%in the regime of $1<m_+/b<3$, in which the zero energy flat bands are expected.\cite{RyuHat02}
Thus, the isolated mid-gap states in Fig. \ref{f:edge_1}(d) 
can be deformed 
into these topologically protected flat bands without the bulk gap closing.
Indeed,  in the phase diagram in Fig. \ref{f:Pha},
the blue and red regions become narrower as $m_-\rightarrow\infty$,
converging respectively to a linear region on $m_+=2$ and $m_+=6$.

What bulk topological invariant
characterizes these weak topological phases embedded in $C=0$?
In contrast to the so-called $\mathbb Z_2$ topological insulator,
the present system lacks time-reversal symmetry.
Yet, as we demonstrate below,
the proposed weak topological phases are protected by
another type of $\mathbb Z_2$ invariant
associated with manifest  reflection symmetry as well as {\it hidden} reflection symmetry.
%
%To reveal symmetry properties of the model, 
To see this,
it is convenient to introduce a unitary-transformed Hamiltonian:
\begin{alignat}1
\widetilde{\cal H} (\bm k) &= U(k_x) {\cal H}(\bm k) U^\dagger(k_x),
\nonumber \\
U(k_x) &= \1_2\otimes {\rm diag} (1,e^{-{\rm i}k_x}),
\end{alignat}
where $\1_2$ operates on the Pauli matrices.
%In the transformed Hamiltonian,
%off-diagonal blocks of $h (\bm k)$ given in Eq. (\ref{hk}) 
%become
%$e^{ik_x} \Gamma_x + e^{-ik_x} \Gamma_x^\dagger$. 
The  transformed Hamiltonian $\widetilde{\cal H}(\bm k)$ is represented simply as
%can be represented as
\begin{alignat}1
\widetilde{\cal H} (\bm k) = \alpha_\mu\gamma_\mu+\beta_\mu\gamma_{2\mu},
\label{h_ab}%-------
\end{alignat}
where
$\gamma$-matrices are defined by 
$\gamma_1=\sigma_x\otimes\sigma_x$,
$\gamma_2=\sigma_y\otimes\1_2$,
$\gamma_3=\sigma_z\otimes\1_2$,
$\gamma_4=\sigma_x\otimes\sigma_y$,
$\gamma_5=\sigma_x\otimes\sigma_z$, 
and $\gamma_{\mu\nu}\equiv {\rm i}[\gamma_\mu,\gamma_\nu]/2$.
The coefficients $\alpha_\mu$ and $\beta_\mu$ are listed in Table \ref{t:Coe}.

%%%%%%%%%%%%%%%%%%%%%%%%%
\begin{table}
\caption{Non-zero coefficients of the Hamiltonian (\ref{h_ab}).}
\label{t:Coe}%-------
\begin{tabular}{llll}
\hline\hline
$\alpha_1$ & $t \sin k_x$ \hfill & $\beta_1$ & $b\cos k_x$\\
$\alpha_2$ & $t \sin k_y$ & $\beta_5$ & $\delta m$\\
$\alpha_3$ & $\left[m+b(\cos k_y-2)\right]$ & &\\
\hline\hline
\end{tabular}
\end{table}
%%%%%%%%%%%%%%%%%%%%%%%%%

We begin by demonstrating the following two properties:
%that 
(i) $\widetilde{\cal H} (\bm k)$ possesses not only  particle-hole symmetry but also
reflection (inversion) symmetry and 
(ii) $\widetilde{\cal H} (\bm k)$ is not periodic with respect to $k_x$,
$\widetilde{\cal H}(k_x+\pi,k_y)=U(\pi)\widetilde{\cal H}(k_x,k_y)U^\dagger(\pi)$.
Here, the form of $U(\pi)= \1_2 \otimes \sigma_z$ implies that 
a twisted boundary condition is imposed on the Hamiltonian $\widetilde{\cal H}(\bm k)$.
Nevertheless,
the half-filled ground states of $\widetilde{\cal H} (\bm k)$ and ${\cal H} (\bm k)$
give the same Chern number on the same Brillouin zone $[0,\pi]\otimes[0,2\pi]$.

(i) Let us first note that 
the Hamiltonian (\ref{h_ab}) has
particle-hole symmetry,
\begin{alignat}1
\Xi\widetilde{\cal H}(\bm k)\Xi^{-1}=-\widetilde{\cal H}(-\bm k),
\label{ParHol}%-------
\end{alignat}
where $\Xi=-{\rm i}\gamma_2\gamma_3K=\sigma_1\otimes\1_2K$ and $K$ is the complex conjugation operator.
The presence of this symmetry is rather natural if one recalls that
eq. (1) is a straightforward extension of the Wilson-Dirac Hamiltonian.
The model has other symmetries described as
\begin{alignat}1
&
P_y\widetilde{\cal H}(k_x,k_y)P_y^{-1}=\widetilde{\cal H}(-k_x,k_y),
\label{RefY}\\ %----------
&
P_x\widetilde{\cal H}(k_x,k_y)P_x^{-1}=\widetilde{\cal H}(k_x,-k_y),
\label{RefX}%--------
\end{alignat}
where $P_y=\gamma_3K$ and $P_x=K$.
These may be regarded as (anti-unitary) reflection symmetry with respect to the $y$- and $x$-directions,
respectively,
and therefore, the model has the inversion symmetry 
$P\widetilde{\cal H}(\bm k)P^{-1}=\widetilde{\cal H}(-\bm k)$,
where $P=P_xP_y=\gamma_3$. 
%The original Hamiltonian ${\cal H} (\bm k)$ has the same 
The symmetries (\ref{ParHol}) and (\ref{RefY}) are also manifested in ${\cal H}(\bm k)$, 
but the symmetry (\ref{RefX}) is hidden in ${\cal H} (\bm k)$.
%, or in other words, it is described 
%by a $k_x$-dependent transformation by the use of $K$ and $U(k_x)$.

(ii) We introduce the Chern number $\widetilde C$ for the transformed Hamiltonian.
Let $\widetilde \psi(k)$ be the negative energy multiplet of the Hamiltonian $\widetilde{\cal H}(\bm k)$
with a phase convention,
\begin{alignat}1
&
\widetilde\psi(-k_x,k_y)=P_y\widetilde\psi(k_x,k_y),
\nonumber\\
&
\widetilde\psi(k_x,-k_y)=P_x\widetilde\psi(k_x,k_y).
\label{PhaCon}%-------------
\end{alignat}
Note that the periodicity of the wave functions is such that 
$\widetilde\psi(k_x+\pi,k_y)=U(\pi)\widetilde\psi(k_x,k_y)$,
where $U(\pi)=-{\rm i}\gamma_1\gamma_4$.
Let
$\widetilde A_i=\widetilde\psi^\dagger \partial_{i}\widetilde\psi$ and 
$\widetilde F_{12}=\epsilon_{ij}{\rm tr}\partial_{i}\widetilde A_j$ 
be the Berry connection and curvature, respectively, where $\partial_{i}\equiv\partial_{k_i}$.
Because of eq. (\ref{PhaCon}), these obey
\begin{alignat}1
&
\widetilde A_j(-k_x,k_y)=(-)^{j-1}\widetilde A_j(k_x,k_y),
\nonumber\\
&
\widetilde A_j(k_x,-k_y)=(-)^{j}\widetilde A_j(k_x,k_y),
\nonumber\\
&
\widetilde F_{12}(-k_x,k_y)=\widetilde F_{12}(k_x,-k_y)=\widetilde F_{12}(k_x,k_y) .
\label{Ber}%-------------
\end{alignat}
The Chern number $\widetilde C$ of the half-filled states
is given by the integration of $\widetilde F_{12}(k)$ 
over the Brillouin zone.
Let us verify $\widetilde C=C$, 
where $C$ is defined in terms of the wave function $\psi (\bm k)$ of the original
Hamiltonian ${\cal H} (\bm k)$ in eq. (\ref{HamK}).
%is the one used in the numerical simulations for Fig. \ref{f:Pha}.
%${\cal N}$ is 
The two wave functions can be related as
$\widetilde{\psi} (\bm k) = U(k_x) \psi (\bm k)$.
This implies that
$\widetilde A_i=A_i+\psi^\dagger U^\dagger \partial_i U\psi$, and hence,
$\widetilde F_{12}=F_{12}+\epsilon_{ij}\partial_i({\rm tr}P_-U^\dagger \partial_j U)$, where 
$A_i$ and $F_{12}$ are the Berry connection and curvature defined through $\psi (\bm k)$, respectively,
and $P_- (\bm k)$ is the projection operator for the occupied states, 
$P_- (\bm k) = \psi (\bm k) \psi^\dagger (\bm k)$.
Then, since $U^\dagger\partial_1 U=-{\rm i}\1_2\otimes{\rm diag}(0,1)$ is a constant matrix and 
$P_- (\bm k)$ is gauge-invariant as well as periodic on the Brillouin zone, 
%the above additional term in the relation 
the above difference  between $\widetilde F$ and $F$
vanishes if it is integrated over the Brillouin zone owing to the Stokes theorem on the torus. 
Thus, we reach $\widetilde{C}=C$.

On the basis of the properties (i) and (ii), 
let us consider a topological invariant that characterizes the weak topological phase studied so far.
Consider the Berry connection $\widetilde A_i$.
Because of eq. (\ref{PhaCon}), the obstruction due to gauge fixing occurs
mainly along the four symmetry lines $k_x=0,\pi/2$ and $k_y=0,\pi$.
%This implies that i
Namely, if the Berry connection is plotted on the Brillouin zone, 
vortices can appear on these lines. Moreover, they are always paired because of eq. (\ref{Ber}).
%The symmetry properties (\ref{Ber})  tell that 
%such vortices appear as pairs on these lines. 
For example, on the Brillouin zone defined by $[0,\pi]\otimes[0,2\pi]$, 
a pair on the $k_x=0$ line sits on the points
%symmetric with respect to $k_y=0$ and $k_y=\pi$. 
$(0,k_y^\star)$ and $(0,2\pi-k_y^\star)$ with the same vorticity.
Of course, 
at some other points away from these symmetry lines, obstructions can also occur.
In this case, vortices appear ``in quartets'',
which are symmetric with 
respect to the four symmetry lines. These also have the same vorticity.
Thus, we know that an even number of vortices always appear 
as long as the phase convention of eq. (\ref{PhaCon}) is adopted. 
However, there are exceptions. Namely, the four crossing points of the four symmetry lines, i.e.,
$(0,0)$, $(\pi/2,0)$, $(\pi/2,\pi)$, and $(0,\pi)$, which will be referred to as $X_1,\cdots,X_4$
in this order.
On these points, single vortices can appear. These single vortices cannot move away from 
these points even if one makes local gauge transformation, since if they did so, they would need an
odd number of partners, as discussed above.
Indeed, in the phase $C=\pm1$ in Fig. \ref{f:Pha}, an odd number of vortices are
located on these four symmetry points $\{X_i\}$ in all gauges, 
as long as the phase convention of eq. (\ref{PhaCon}) is used. See Table \ref{t:Inv}.

This implies that unpaired vortices on these points 
%carry some important information of 
can be used to reveal the topological 
properties of the present system. Indeed, these vortices inform us of the ``parity'' of the Chern number.
Moreover, different configurations of vortices on $\{X_i\}$ imply topologically different phases,
since the location of these vortices is gauge-invariant. 
Therefore, it is natural to expect that the $C=0$ phase 
can be further distinguished by the obstruction on the four symmetry points.
Note that the present Hamiltonian $\widetilde{\cal H}$ has inversion symmetry. 
Therefore, the obstruction on the four points $\{X_i\}$ is associated with the parity of the wavefunctions.
With respect to the parity operator $P$, eq. (\ref{PhaCon}) implies that we choose  
\begin{alignat}1
\widetilde\psi(-k)=P\widetilde\psi(k) .
\label{GauFixInv}%----------
\end{alignat}
At the point $X_1=(0,0)$, for example, this means that $\widetilde\psi(0)=P\widetilde\psi(0)$. 
%On the other hand,
%since $P^2=\1_{2q}$, each wavefunction of the $q$ multiplet has definite parity $\pm1$.
Let us assume that among the two occupied wavefunctions of $\widetilde\psi(0)$, 
$n$ wavefunctions have a parity of $-1$ and the others have a parity of $+1$.
Then, the former are the obstructions of the gauge-fixing condition (\ref{GauFixInv}).
When $n$ is even ($n=0$ or $2$), there appear an even number of vortices at $X_1$, 
which can move away from this point via
suitable gauge transformation. However, if $n$ is odd ($n=1$), 
%at least (or generically speaking)
one vortex is forced to locate at $X_1$.
The $\mathbb{Z}_2$ invariant in this context can be extracted from
$\det[\widetilde\psi^\dagger(0)P\widetilde\psi(0)]=\pm1$, 
where the case $-1$ is the $\mathbb{Z}_2$ obstruction of the condition
(\ref{GauFixInv}).

%%%%%%%%%%%%%%%%%%%%%%%%%%%%%%%%%%
\begin{table}
\caption{ 
$\mathbb{Z}_2$ invariant numerically computed. The top two cases correspond 
to the uniform mass Wilson-Dirac model
$\delta m=0$ ($b=2$ and $t=1$) in the trivial $C=0$ phase, 
whereas the middle two cases are for the $C=\pm1$ phases.
The bottom two cases are for the weak topological phases A and B.
These six cases are indicated as black points in Fig. \ref{f:Pha}.  
}
\label{t:Inv}%-------
\begin{tabular}{ccccc}
\hline\hline
$m_+$&$m_-$&$C$&$[n_1n_2n_3n_4]$&$[[\tilde{n}_1\tilde{n}_2\tilde{n}_3\tilde{n}_4]]$ \\
\hline
$-1$&$-1$&0&[0110]&[[0000]]\\
9&9&0&[0110]&[[0000]]\\
\hline
0.5&1&1&[1110]&[[1000]]\\
7&8&$-1$&[0111]&[[0001]]\\
\hline
2.5&5.5&0&[1111]&[[1001]]\\
3.5&7&0&[0101]&[[0011]]\\
\hline\hline
\end{tabular}
\end{table}
%%%%%%%%%%%%%%%%%%%%%%%%%%%%%%%%%%%%%

This simple observation is extended to other symmetry points $X_i$. 
Here, it should be
noted that the wavefunction $\widetilde\psi(k)$ is not periodic with respect to $k_x$
because of the twist operator $U(k_x)$, as discussed below eq. (\ref{PhaCon}).
In particular, 
$\widetilde\psi(-k_x+\pi,\bar{k}_y)=U(\pi)\widetilde\psi(-k_x,\bar{k}_y)=U(\pi)P\widetilde\psi(k_x,\bar{k}_y)$,
where $\bar{k}_y=0$ or $\pi$, and the phase convention at $k_x=\pi/2$ is thus modified.
%: Namely, $\widetilde\psi(k_x+\pi,k_y)=U(\pi)\widetilde\psi(k_x,k_y)$.
Note that $P=\sigma_3\otimes\1_2$ and $U(\pi)=\1_2\otimes\sigma_3$ 
and hence $U(\pi)P=\sigma_3\otimes\sigma_3$.
Thus, by taking this periodicity into account, the condition (\ref{GauFixInv}) 
is explicitly written as
\begin{alignat}1
&
\widetilde\psi(X_i)=P_i\widetilde\psi(X_i), \quad i=1, \cdots,4,
\end{alignat}
where $P_i\equiv \sigma_3\otimes\1_2$ ($i=1,4$) and $P_i\equiv\sigma_3\otimes\sigma_3$ ($i=2,3$).
%This implies that at the symmetry points $X_i$, the wavefunctions do not have definite parity
%because of the nontrivial periodicity of the wavefunctions.
This leads to the $\mathbb{Z}_2$ obstruction
\begin{alignat}1
\delta_i=\det \widetilde\psi^\dagger(X_i)P_i\widetilde\psi(X_i),
\label{AnaWeaInv}%-------
\end{alignat}
%Then, 
%\begin{alignat}1
%w_i\equiv \det w(X_i) ,
%\end{alignat}
which can take $\pm1$ only. 
Now let us define a set of four numbers $[n_1n_2n_3n_4]$, where $\delta_i=(-1)^{n_i}$.
These numbers are gauge-invariant modulo 2, indicating that 
$n_i=1$ (0) implies obstruction (no obstruction) at $X_i$.

In Table \ref{t:Inv}, we show several examples of this invariant.
Note that, even in trivial $C=0$ phases on the $m_+ = m_-$ line, 
the obtained invariant is [0110], 
which seems nontrivial at first sight. This is, however, due to 
the effect of the twist operator $U(\pi)$. Generally, $U(\pi)=\1_2\otimes\sigma_3$ implies that 
among four states, half are periodic and the other half are anti-periodic in $k_x$; 
therefore, the half-filled 
ground states include one periodic state and one anti-periodic state.
Therefore, we should define the relative $\mathbb{Z}_2$ invariant
as ``excitations'' for this background obstruction such that 
$[[\tilde{n}_1\tilde{n}_2\tilde{n}_3\tilde{n}_4]]=[n_1n_2n_3n_4]-[0110]$ mod 2.

This is the bulk $\mathbb{Z}_2$ invariant that characterizes the weak topological phases in the 
present superlattice system.
From Table \ref{t:Inv}, we can interpret phase A as $[[1001]]=[[1000]]+[[0001]]$ and thus
the $C=1+(-1)=0$ nature of this phase is established.
It is also clear that the phase B, assigned $[[0011]]$, is distinguished from this phase. 
%originated from flat bands owing to
%chiral symmetry in the $m_-\rightarrow\infty$.
%These two nontrivial phases are topologically distinct phases ensured by the $\mathbb{Z}_2$ invariant
%$[[\tilde{n}_1\tilde{n}_2\tilde{n}_3\tilde{n}_4]]$.

Let us finally mention how
the $\mathbb{Z}_2$ indices $\tilde{n}_i$ 
%at the four symmetry points $X_i$
%that appear
%in the representation $[[\tilde{n}_1\tilde{n}_2\tilde{n}_3\tilde{n}_4]]$
are related to the edge spectrum
in the two different ribbon geometries  shown in Fig. \ref{f:edge_1}.
Define $\tilde \delta_i=(-1)^{\tilde n_i}$.
This is the relative $\delta_i$ in eq. (\ref{AnaWeaInv}) with respect to the background.
%Since $\tilde{n}_i$ is either 0 or 1,
%the quantity $\delta_i=(-1)^{\tilde{n}_i}$ takes either $+1$ or $-1$.
Then, as in ref. \cite{FuKane},
we consider the product of two $\tilde\delta_i$'s [$\tilde\delta_{\bar{k}_\mu 1}$ and $\tilde\delta_{\bar{k}_\mu 2}$]
on the line $k_\mu = \bar{k}_\mu$ ($\bar{k}_y=0, \pi$, while $\bar{k}_x=0, \pi/2$)
and define the quantity
$\pi_{\bar{k}_\mu} = \tilde\delta_{\bar{k}_\mu 1} \tilde\delta_{\bar{k}_\mu 2}$.
Again, $\pi_{\bar{k}_\mu}=\pm 1$.
If $\pi_{\bar{k}_\mu}=-1$,
the edge spectrum in the ribbon geometry directed to the $\mu$-axis 
becomes gapless at $k_\mu=\bar{k}_\mu$;
otherwise, the spectrum is gapped.
Here, this statement can be verified empirically using the explicit values of 
$\tilde{n}_i$ listed in Table II,
while the proof of the statement involves the calculation of the Berry curvature 
integrated along the loop $k_\mu=\bar{k}_\mu$ (Wilson loop).
We leave detailed description of this proof to a forthcoming publication.

\section*{Acknowledgments}
We are supported by KAKENHI.
T.F. and K.I. are supported by grants-in-aid for the ``Topological Quantum Phenomena''
(Nos. 23103502 and 23103511), T.F. by Grant No. 25400388, 
and Y.H. by Grant Nos. 23340112, 23654128, 23540460, and 25610101.

\end{document}